\documentclass[prl,aps,twocolumn,superscriptaddress,showpacs]{revtex4}

\usepackage{graphicx}

\usepackage{amsmath}
\usepackage{amsfonts}

\newcommand{\abs}[1]{\left|#1\right|}

\newcommand{\rmi}{{\rm i}}
\newcommand{\rmd}{{\rm d}}

\newcommand{\ket}[1]{|#1\rangle}
\newcommand{\beq}{\begin{equation}}
\newcommand{\eeq}{\end{equation}}

\begin{document}
\title{A Single Atom Transistor in a 1D Optical Lattice}
\author{A. Micheli}
\author{A. J. Daley}
\affiliation{Institute for Quantum Optics and Quantum Information of the Austrian Academy
of Sciences, A-6020 Innsbruck, Austria} \affiliation{Institute for Theoretical Physics,
University of Innsbruck, A-6020 Innsbruck, Austria}
\author{D. Jaksch}
\affiliation{Clarendon Laboratory, University of Oxford, Parks Road, Oxford OX1 3PU,
U.K.}
\author{P. Zoller}
\affiliation{Institute for Quantum Optics and Quantum Information of the Austrian Academy
of Sciences, A-6020 Innsbruck, Austria} \affiliation{Institute for Theoretical Physics,
University of Innsbruck, A-6020 Innsbruck, Austria}
\date{June 3, 2004}

\begin{abstract}
We propose a scheme utilising a quantum interference phenomenon to
switch the transport of atoms in a 1D optical lattice through a
site containing an impurity atom. The impurity represents a qubit
which in one spin state is transparent to the probe atoms, but in
the other acts as a single atom mirror. This allows a single-shot
quantum non-demolition measurement of the qubit spin.
\end{abstract}

\maketitle

\begin{figure}
\begin{center}
\includegraphics[width=0.48\textwidth]{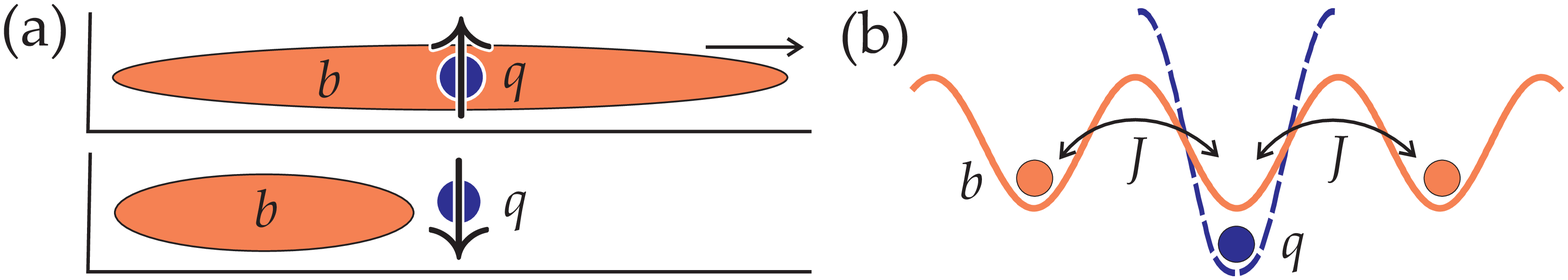}
\caption{(a) A spin 1/2 impurity used as a switch: in one spin state it is transparent to
the probe atoms, but in the other it acts as a single atom mirror. (b) Implementation of
the SAT as a separately trapped impurity $q$ with probe atoms $b$ in an optical
lattice.}\label{Fig:setup}
\end{center}
\end{figure}

Coupling of a spin 1/2 system to Bosonic and Fermionic modes is
one of the fundamental building blocks of quantum optics and solid
state physics. Motivated by the recent progress with cold atoms in
1D \cite{lattice1d}, we consider a spin 1/2 atomic impurity which
is used to switch the transport of either a 1D Bose-Einstein
Condensate (BEC) or a 1D degenerate Fermi gas initially situated
to one side of the impurity. In one spin state the impurity is
transparent to the probe atoms, whilst in the other it acts as
single atom mirror, prohibiting transport via a quantum
interference mechanism reminiscent of electromagnetically induced
transparency (EIT) \cite{eit} (Fig.~\ref{Fig:setup}a). Observation
of the atomic current passing the impurity can then be used as a
quantum non-demolition (QND) measurement \cite{qnd} of its
internal state, which can be seen to encode a qubit,
$\ket{\psi_q}=\alpha\ket{\!\!\uparrow}+\beta\ket{\!\!\downarrow}$.
If a macroscopic number of atoms pass the impurity, then the
system will be in a macroscopic superposition,
$\ket{\Psi(t)}=\alpha\ket{\!\!\uparrow}\ket{\phi_\uparrow(t)}+\beta\ket{\!\!\downarrow}\ket{\phi_\downarrow(t)}$,
which can form the basis for a single shot readout of the qubit
spin. Here, $\ket{\phi_{\sigma}(t)}$ denotes the state of the
probe atoms after evolution to time $t$, given that the qubit is
in state $\sigma$ (Fig.~\ref{Fig:setup}a). In view of the analogy
between state amplification via this type of blocking mechanism
and readout with single electron transistors (SET) used in solid
state systems \cite{set}, we refer to this setup as a Single Atom
Transistor (SAT).

We propose the implementation of a SAT using cold atoms in 1D optical lattices
\cite{oltheory,meltingmi,greiner,sdol}. We consider probe atoms $b$ to be loaded in the
lattice to the left of a site containing the impurity atom, which is trapped by a
separate (e.g., spin-dependent \cite{sdol}) potential (Fig.~\ref{Fig:setup}b). The
passage of $b$ atoms past the impurity $q$ is then governed by the spin-dependent
effective collisional interaction $H_{\rm int}=\sum_\sigma U_{\rm
eff,\sigma}\hat{b}_0^\dag \hat{b}_0 \hat{q}_\sigma^\dag \hat{q}_\sigma$. By making use of
a quantum interference mechanism, we engineer complete complete blocking (effectively
$U_{\rm eff}\rightarrow \infty$) for one spin state and complete transmission ($U_{\rm
eff}\rightarrow 0$) for the other. Below we first consider the detailed scattering
processes involved in the transport of a single particle through the SAT, and then
generalise this to interacting many-particle systems including a 1D Tonks gas.

\begin{figure}
\begin{center}
\includegraphics[width=0.48\textwidth]{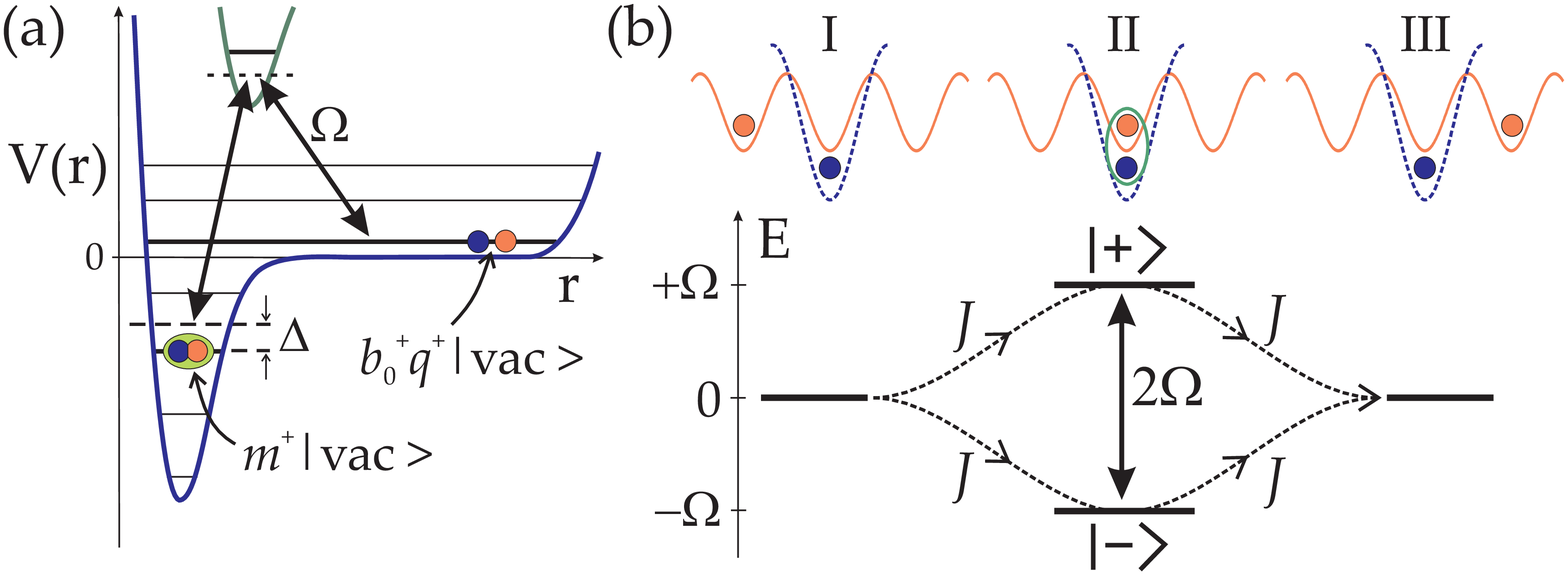}
\caption{(a) The optical Feshbach setup couples the atomic state
$\hat{b}^\dag_0\hat{q}^\dag_\sigma \ket{{\rm vac}}$ (in a particular motional state
quantised by the trap) to a molecular bound state of the Born-Oppenheimer potential,
$m^\dag_\sigma\ket{{\rm vac}}$, with effective Rabi frequency $\Omega_\sigma$ and
detuning $\Delta_\sigma$. (b) A single atom passes the impurity (I$\rightarrow$III) via
the two dressed states (II), $\ket{+}=\hat{b}^\dag_0\hat{q}^\dag_\sigma \ket{{\rm vac}}
+m^\dag_\sigma\ket{{\rm vac}}$ and $\ket{-}=\hat{b}^\dag_0\hat{q}^\dag_\sigma \ket{{\rm
vac}}-m^\dag_\sigma\ket{{\rm vac}}$. Quantum interference between the paths gives rise to
an effective tunnelling rate $J_{\rm eff,\sigma}$.}\label{Fig:EIT}
\end{center}
\end{figure}

The quantum interference mechanism needed to engineer $U_{\rm
eff}$ can be produced using an optical or magnetic Feshbach
resonance \cite{feshbach}. For the optical case a Raman laser
drives a transition on the impurity site, $0$, from the atomic
state $\hat{b}^\dag_0 \hat{q}^\dag_\sigma \ket{{\rm vac}}$ via an
off-resonant excited molecular state to a bound molecular state
back in the lowest electronic manifold
$\hat{m}^\dag_\sigma\ket{{\rm vac}}$ (Fig.~\ref{Fig:EIT}a). We
denote the effective two-photon Rabi frequency and detuning by
$\Omega_\sigma$ and $\Delta_\sigma$ respectively. The Hamiltonian
for our system is then given ($\hbar\equiv 1$) by
$\hat{H}=\hat{H}_b+\hat{H}_0$, with
\begin{eqnarray}
\hat{H}_{\rm b}&=&-J\sum_{\langle ij \rangle} {\hat b}_i^\dag {\hat b}_j + \frac{1}{2}
U_{bb} \sum_j
{\hat b}_j^{\dag}{\hat b}_j\left({\hat b}_j^{\dag}{\hat b}_j-1 \right) \nonumber \\
\hat{H}_{\rm 0}&=&\sum_\sigma\left[ \Omega_\sigma \left({\hat m}_\sigma^\dag {\hat
q}_\sigma {\hat b}_0 + {\rm h.c}\right) - \Delta_\sigma {\hat m}_\sigma^\dag {\hat
m}_\sigma \right]
\nonumber\\
&+& \sum_\sigma \left[U_{qb,\sigma} {\hat b}_0^\dag {\hat q}_\sigma^\dag {\hat q}_\sigma
{\hat b}_0 + U_{bm,\sigma} {\hat b}_0^\dag {\hat m}_\sigma^\dag {\hat m}_\sigma {\hat
b}_0\right],\label{Eq:Hamiltonian}
\end{eqnarray}
where the operators $\hat{b}$ obey the standard commutation
(anti-commutation) relations for Bosons (Fermions). $H_{\rm b}$
gives a Hubbard Hamiltonian for the $b$ atoms with tunnelling
matrix elements $J$ giving rise to a single Bloch band with
dispersion relation $\varepsilon(k)=-2J\cos k a$ ($a$ is the
lattice spacing), and collisional interactions (which are non-zero
only for Bosons) given by $U_{bb}=4\pi\hbar^2 a_{bb}\int
d^3\mathbf{x} \abs{{\rm w}_j(\mathbf{x})}^4/m_b$, where ${\rm
w}_j(\mathbf{x})$ is the Wannier-function for a particle localized
on site $j$, $a_{bb}$ is the scattering length for $b$ atoms and
$m_b$ is their mass. $H_0$ describes the additional dynamics due
to the impurity on site 0, where atoms $b$ and $q$ are converted
to a molecular state with effective Rabi frequency $\Omega_\sigma$
and detuning $\Delta_\sigma$, and the last two terms describe
background interactions, $U_{\alpha\beta,\sigma}$ for two
particles $\alpha, \beta \in \{q_\sigma,b,m\}$, which are
typically weak. This model is valid for
$U_{\alpha\beta},J,\Omega,\Delta \ll \omega$, where $\omega$ is
the energy separation between Bloch bands. Because the dynamics
for the two spin channels $q_\sigma$ can be treated independently,
in the following we will consider a single spin channel, and drop
the subscript $\sigma$.

For off-resonant laser driving ($\Omega \ll \abs{\Delta}$), the Feshbach resonance
enhances the interaction between $b$ and $q$ atoms, giving the familiar result $U_{\rm
eff} = U_{qb}+\Omega^2/\Delta$. However, for resonant driving ($\Delta=0$) the physical
mechanism changes, and the effective tunnelling $J_{\rm eff}$ of an atom $b$ past the
impurity (Fig.~\ref{Fig:EIT}b, $\rm I \rightarrow III$) is blocked by quantum
interference. On the impurity site, laser driving mixes the states
$\hat{b}^\dag_0\hat{q}^\dag \ket{{\rm vac}}$ and $m^\dag\ket{{\rm vac}}$, forming two
dressed states with energies $\varepsilon_\pm=(U_{qb})/2 \pm(U_{qb}^2/4+\Omega^2)^{1/2}$
(Fig.~\ref{Fig:EIT}b, II). The two resulting paths for a particle of energy $\varepsilon$
destructively interfere so that for large $\Omega \gg J$ and $U_{qb}=0$, $J_{\rm
eff}=-J^2/(\varepsilon + \Omega)-J^2/(\varepsilon - \Omega)\rightarrow 0$. This is
analogous to the interference effect underlying EIT \cite{eit}, and is equivalent to
having an effective interaction $U_{\rm eff}\rightarrow \infty$. In addition, if we
choose $\Delta=-\Omega^2/U_{qb}$, the paths constructively interfere, screening the
background interactions to produce perfect transmission ($U_{\rm eff}\rightarrow 0$).

\begin{figure}[ptb]
\begin{center}
\includegraphics[width=0.48\textwidth]{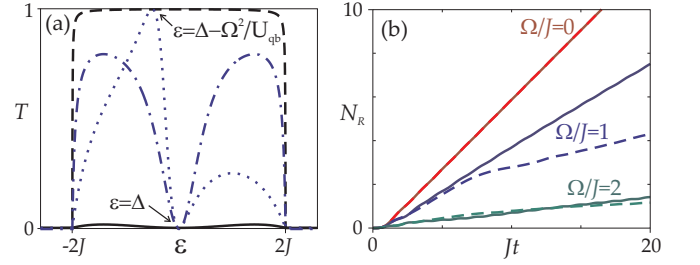}
\caption{(a) SAT transmission coefficients $T\equiv|f^{(+)}|^2$ for a particle $b$ as a
function of its energy $\varepsilon(k)$ for $\Omega/J=4,\Delta=0,U_{qb}/J=0$ (solid
line), $\Omega/J=8,\Delta/J=4,U_{qb}/J=2$ (dashed line), $\Omega/J=1,\Delta=0,U_{qb}/J=2$
(dotted), and $\Omega/J=1,\Delta=0,U_{qb}/J=0$ (dash-dot). (b) The number of particles to
the right of the impurity, $N_R(t)$, from exact numerical calculations for Bosons in the
limit $U_{bb}/J\rightarrow\infty$ (dashed lines) and Fermions (solid lines) in a 1D Mott
Insulator state with $n=1$, for $\Delta=0$, $\Omega/J=0,1,2$.} \label{Fig:transmit}
\end{center}
\end{figure}

For a more detailed analysis, we solve the Lippmann-Schwinger equation exactly for
scattering from the impurity of an atom $b$ with incident momentum $k>0$ in the lowest
Bloch-band. The resulting forwards and backwards scattering amplitudes, $f^{(\pm)}(k)$
respectively, are
\begin{equation}
f^{(\pm)}(k) = \left[1+\left(\frac{\rmi a U_{\rm eff}(k)}{v(k)}\right)^{\pm
1}\right]^{-1},
\end{equation}
where the energy dependent interaction $U_{\rm
eff}=U_{qb}+\Omega^2/(\varepsilon(k)-\Delta)$ and the
phase-velocity $v(k)=\partial \varepsilon/\partial k =2Ja\sin k
a$. The corresponding transmission probabilities, $T(k)=
\abs{f^{(+)}(k)}^2$, are plotted in Fig.~\ref{Fig:transmit}a as a
function of $\varepsilon(k)$ for various $\Omega$ and $\Delta$.
For intermediate $\Omega\sim J$, these are Fano-profiles with
complete reflection at $\varepsilon(k)=\Delta$ and complete
transmission at $\varepsilon(k)=\Delta-\Omega^2/U_{qb}$. The SAT
thus acts as an energy filter, which is widely tunable via the
laser strength and detuning used in the optical Feshbach setup.
For $\Omega>4J$, $T$ is approximately independent of $k$, and we
recover the previous result, i.e., that transport can be
completely blocked or permitted by appropriate selection of
$\Delta$. Note that this mechanism survives the generalisation to
a multi-band Hubbard model, where higher energy Bloch bands are
included in the calculation, and is resistant to loss process,
which are discussed below.

We now consider the full many-body dynamics of $N$ probe atoms $b$
initially prepared in the ground state in a trap (box) of $M$
lattice sites on the left side of the impurity $q$. We are then
interested in the expectation value of the steady state coherent
current ${\hat I}=d{\hat N}_R/dt$ (where ${\hat N}_R =
\sum_{j>0}{\hat b}_j^\dag {\hat b}_j$ is the number of particles
on the right side of the impurity, see Fig.~\ref{Fig:transmit}b),
which depends on the laser parameters, the filling factor,
$n=N/M$, and, for Bosons, the interaction strength, $U_{bb}$. We
first consider the case of a dilute or noninteracting gas, before
treating both interacting Bosons, and non-interacting Fermions
with arbitrary $n$.

For a dilute noninteracting Bose quasi-condensate ($n\ll 1, U_{bb}=0$), or for any very
dilute gas, (where the momentum distribution is very narrow), the behaviour is very
similar to that of a single particle. If the gas is quickly accelerated to a finite
momentum $k$, e.g., by briefly tilting the lattice, then the atoms will coherently tunnel
through the impurity according to the scattering amplitudes $f^{(\pm)}(k)$. The resulting
current $I\propto N \abs{f^{(+)}(k)}^2 v(k)$, where n is the initial filling factor on
the left of the impurity, and $v(k)$ is the velocity of a Bloch-wave with momentum $k$.

\begin{figure}[ptb]
\begin{center}
\includegraphics[width=0.48\textwidth]{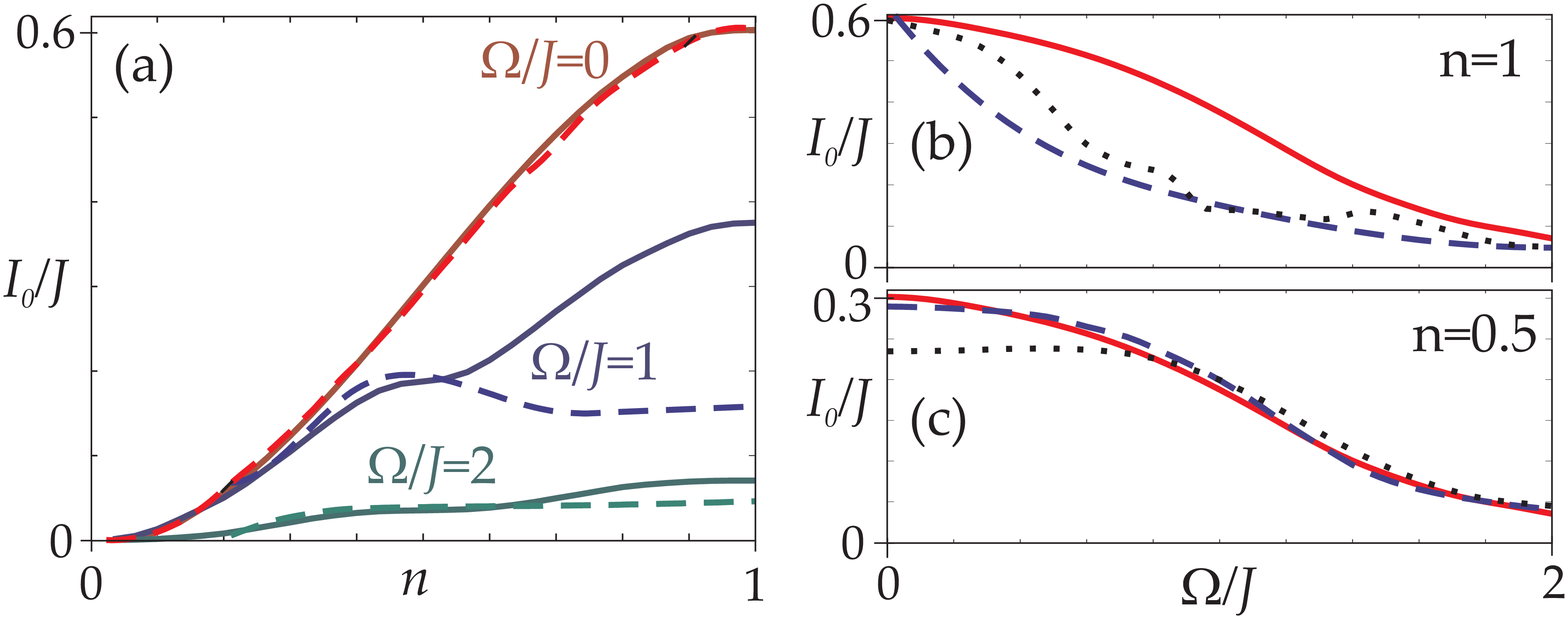}
\caption{(a) The steady state current of $b$ atoms through the impurity as function of
the initial filling factor $n$ for $\Delta=0$ and $\Omega/J=0,1,2$. The solid lines show
the analytic result $I_0$ for Fermions, whereas the dashed lines show the exact numerical
result for hard-core Bosons with $U_{bb}/J\rightarrow\infty$. For $\Omega=0$ these
results are indistinguishable. (b, c) The steady state current as function of the
Rabi-frequency $\Omega/J$ on resonance $\Delta=0$ for (b) unit filling and (c) half
filling. The solid lines show the analytic result for Fermions, whereas the dashed
(dotted) lines give numerical results for Bosons with $U_{bb}/J=20$ ($4$).}
\label{Fig:DCCurrents}
\end{center}
\end{figure}

For a Fermi gas the equations of motion are linear and may be solved exactly provided
$U_{bm}= U_{qb}$. Scattering from the impurity then occurs independently for each
particle in the initial Fermi sea, with scattering amplitudes $f^{(\pm)}(k)$ for $k\leq
k_F$, where the Fermi momentum $k_F=\pi n$. After a short transient period, on the order
of the inverse tunnelling rate $1/J$, the system establishes a roughly constant flux of
particles through the impurity (Fig.~\ref{Fig:transmit}b), with a time-averaged current
for resonant driving $\Delta=0$ given by
\begin{eqnarray}
I_0 &=& \frac{1}{\pi a}\int_{-2J}^{-2J+\varepsilon_F} \rmd \varepsilon f(\varepsilon) T(\varepsilon)v(\epsilon)\label{Eq:Current_Resonant_Mott}\\
&=& \frac{J}{\pi}\left[V-\frac{G_+{\rm arctan}\frac{VG_-}{G_+^2-V}-G_-{\rm
arctanh}\frac{VG_+}{G_-^2+V}} {\frac{G_+^2+G_-^2}{G_+G_-}}\right],\nonumber
\end{eqnarray}
where $\varepsilon_F$ is the Fermi Energy in the initial state, $f(\varepsilon)$ is the
density of states per site (on the left of the impurity),
$2G_\pm^2\equiv(1+\Omega^4/4J^4)^{1/2}\pm 1$ and $V\equiv\varepsilon_F/2J=\sin^2(n
\pi/2)$.

For a Tonks gas of strongly interacting Bosons ($U_{bb}/J\gg 1$ with $n \leq 1$) we
expect to observe similar behaviour to that observed for Fermions. In this limit, double
occupation of a site can be neglected, and the behaviour can be mapped onto Fermionic
particles via a Jordan-Wigner transformation \cite{jordanwigner}. The Hamiltonian is then
the same up to a nonlinear phase factor $\Omega\rightarrow \Omega (-1)^{{\hat N}_R}$,
which essentially causes $\Omega$ to change sign when a particle passes the impurity. The
contribution of this phase factor should be small for weak coupling, $\Omega\ll J$, and
also for strong coupling, where no particles will tunnel through the impurity, i.e.,
$N_R\simeq 0$.

For the general case of many Bosons we perform exact numerical integration of the time
dependent Schr\"{o}dinger equation for the Hamiltonian \eqref{Eq:Hamiltonian} using
Vidal's algorithm for ``slightly entangled quantum states''\cite{vidal}. This algorithm
selects adaptively a decimated Hilbert space on which a state is represented, by
retaining at each time step only those basis states that carry the greatest weight in
Schmidt decompositions taken from bipartite splittings of the system between each two
lattice sites. A sufficiently large decimated Hilbert space is then selected so that the
results of the simulations are essentially exact. For each set of parameters we first
prepared the initial state via an imaginary time evolution which found the ground state
for atoms in a box trap on the left of the impurity. Then, considering initially a single
impurity atom $q$ on the site $0$ and unoccupied sites to the right of that site, we
calculated the time evolution of the system until it had reached a quasi-steady state
behaviour. These simulations allowed us to obtain the behavior for finite repulsion
$U_{bb}$, and also to test the effects of the nonlinear phase factor $(-1)^{{\hat N}_R}$
for strongly interacting Bosons, $U_{bb}\rightarrow \infty$.

In Fig.~\ref{Fig:transmit}b we plot the number of particles on the right of the impurity
$N_R(t)$ for Fermions and for Bosons with $U_{bb}/J\rightarrow\infty$, starting from a
Mott Insulator (MI) state with $n=1$, for $\Delta=0$, $\Omega/J=0,1,2$. For $\Omega=0$
the results for Bosons and Fermions are identical, whilst for $\Omega/J=1,2$, we observe
an initial period for the Bosons in which the current is similar to that for the
Fermionic systems, after which the Bosons settle into a steady state with a significantly
smaller current. The initial transient period for the Bosons incorporates the settling to
steady state of firstly the molecule dynamics, and secondly the momentum distribution on
the right of the impurity. These transients are suppressed if $\Omega$ is ramped slowly
to its final value from a large value $\Omega > 4J$.

The dependence of the steady state current on the initial filling factor $n$ is depicted
in Fig.~\ref{Fig:DCCurrents}a for resonant driving with $\Omega/J=0,1,2$. For
$\Omega=U_{qb}=U_{bm}=0$, the current $I_0=2J\sin^2(n\pi/2)/\pi$ is identical for
Fermions and hard-core Bosons ($U_{bb}\rightarrow \infty$), as we expect from the exact
correspondence given in this limit by the Jordan-Wigner transformation. For Fermions with
weak, resonant laser driving, the main features of the Fano profile
(Fig.~\ref{Fig:transmit}a) are observed in correspondence with the integral in
(\ref{Eq:Current_Resonant_Mott}). For example, a plateau in $I_0(n)$ is observed near
$n=\arccos(-\Delta/2J)/\pi=1/2$, as the Fermi Energy is raised past
$\varepsilon\sim\Delta=0$, which corresponds to the zero of the transmission probability
$T(\varepsilon)$. Good agreement is also observed with the result for Bosons in this
limit with $n<1/2$, whilst for larger $n$ Bosons are blocked better, with a factor of
$2-3$ in the steady state currents.

The enhanced blocking for Bosons is also seen in Fig.~\ref{Fig:DCCurrents}b, where we
plot the steady state current as a function of $\Omega$ for resonant driving and $n=1$.
It is clear from these figures together that this difference is a feature of the
parameter regime $n>1/2$, $\Omega\sim J$, which can be linked directly to phase factor of
$(-1)^{N_R}$ that arises in the Jordan-Wigner transformation. As $\Omega$ is increased
and fewer particles pass the impurity, the results for Fermions and Bosons again converge
as expected. For small $\Omega$ there are small differences between Bosons with finite
$U_{bb}/J=4$ and ($U_{bb}/J\rightarrow\infty$), with the currents always lower than the
equivalent fermionic current, owing largely to the smaller mean squared momentum in the
initial state. For large driving, $\Omega\gg 4 J$ the basic interference process is
extremely efficient for both Bosons and Fermions, and we observe complete blocking or
transmission of the current by quantum interference for the appropriate choice of
$\Delta$.

\begin{figure}
\begin{center}
\includegraphics[width=0.48\textwidth]{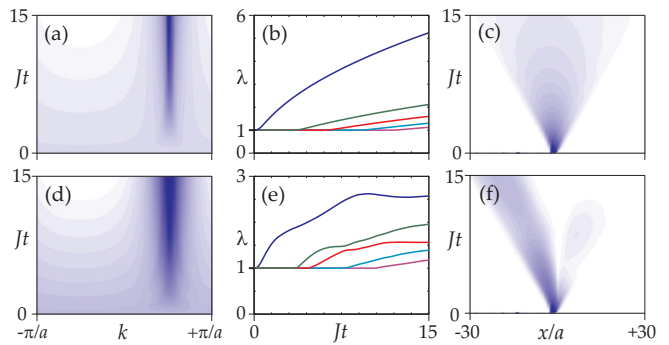}
\caption{Exact numerical results showing the propagation of $N=30$
hard-core Bosons in an initial MI state ($n=1$) through a SAT with
(a-c) $\Omega=0$ and (d-f) $\Omega=0.5 J$, with
$\Delta=U_{qb}=U_{bm}=0$. The plot shows (a,d) the momentum
distribution, (b,e) the five largest eigenvalues, $\lambda_m$, of
the single particle density matrix, $\langle
\hat{b}_i^\dag\hat{b}_j\rangle$ and (c,f) the spatial density of
the largest eigenmode (the quasi-condensate) as a function of
time. Darker colours represent higher values.}\label{Fig:timedep}
\end{center}
\end{figure}

In Fig.~\ref{Fig:timedep} we investigate the time evolution of $30$ hard-core Bosons
($U_{bb}\rightarrow \infty$) in an initial MI state, which are released through a SAT
which is switched at $t=0$. For $\Omega=0$, we see that as the gas expands the momentum
distribution becomes peaked as a quasi-condensate is formed with $k=\pi/2a$, which
consists of a coherent superposition of particles propagating to the right and holes
propagating to the left as the MI state melts \cite{meltingmi}. This mode grows outwards
from the edge of the initial distribution, and contains at its peak $\sim \sqrt{N}$
particles, as is expected for such dynamically formed quasi-condensates in a 1D lattice
\cite{muramatsu}. In contrast, for $\Omega/J=0.5$, the momentum distribution is broader,
and the quasi-condensate mode contains many fewer particles. The mode also consists of
distinct branches, holes in the melting MI propagating to the left and particles to the
right, which are initially coherent, but become decoupled at $t\sim 12/J$. For larger
$\Omega$ this behaviour becomes more pronounced, and for $\Omega>4J$, the MI state
essentially remains unchanged.

The melting of a MI in this way can be used as the basis for a convenient single-shot
measurement of the spin state of $q$. If $q$ is in a superposition of spin states, only
one of which will permit transport of the $b$ atoms, then after some propagation time,
the system will be in a macroscopic superposition of distinct quantum phases (MI and
quasi-condensates). These are distinguishable because if the $b$ atoms are released from
the lattice, the quasi-condensate will produce an interference pattern, whereas the MI
state will not. The visibility of the resulting pattern can thus be used to measure the
qubit spin.

A remarkable feature of the SAT is its resistance to both two- and
three-body loss processes induced by the optical Feshbach
resonance on the impurity site. Spontaneous emissions from the
off-resonant excited molecular state amount to a two-body loss
process at a rate $\sim \gamma_{2B}$ in the states $\ket{+}$ and
$\ket{-}$. These small rates are further suppressed in the
blocking regime $J,\gamma_{2B}\ll\Omega$, with the resulting
decoherence rate $\gamma_{\rm dec}\propto J^2\gamma_{2B}n
/\Omega^2$, with $n$ the mean site occupation of the $b$ atoms.
Collisions of atoms $b$ with molecules $m$ \cite{petrov3body} are
strongly suppressed in the Tonks gas regime, as well as for
Fermions. For a weakly interacting Bose gas the corresponding
three-body loss rate, $\gamma_{3B}$, is again strongly suppressed
in the blocking regime ($J,\gamma_{3B}\ll \Omega$) with
$\gamma_{\rm dec}\propto J^4\gamma_{3B}n^2 /\Omega^4$.

Parallels may be drawn between the SAT and other systems coupled
to fermionic and bosonic modes. These include the QND-readout of a
single photon in cavity-quantum electrodynamics \cite{cqnd},
electron counting statistics \cite{levitov}, and the transport of
electrons past impurities such as quantum dots \cite{cazmas}
(although there particles are normally initially present on both
sides of the impurity). However, the long decoherence times for
atoms in optical lattices imply coherent transport over longer
timescales than is observed in these other systems, which are
inherently dissipative. In addition, blocking and/or energy
filtering by one or more SATs could be applied as tools in the
study of Bose and Fermi gases in a 1D lattice.

Work in Innsbruck is supported by EU Networks and the Institute for Quantum Information.
DJ is supported by the IRC on quantum information processing.

\end{document}